# About possibility of strong coupling between single molecule and surface plasmons

Mykhaylo M. Dvoynenko

*Lashkaryov's Institute of Semiconductor Physics, National Academy of Sciences of Ukraine 41, pr. Nauki, Kyiv, 03028, Ukraine*

**Abstract:**

The possibility of strong coupling between a single molecule and surface plasmons is analyzed on the basis of a microscopic classical description. It is predicted that strong single molecule – plasmon coupling can happen for a silver nanoparticle only in the UV range where the real part of the dielectric function of silver approaches a value of -1. A critical view is taken on an interpretation of spectroscopic data of composite layers of dye molecules and a metal film in terms of single molecule-plasmon coupling in a Kretschman configuration in some recent publications. It is shown that experimental results interpreted as a mixed (hybrid) radiation do not reveal strong coupling and can be explained without any coupling effects.

A fundamental phenomenon in physics is the strong coupling in light-matter interactions [1]. Initially strong coupling leading to a splitting (called vacuum Rabi splitting) in fluorescence [2] and absorption [3] spectra was associated with an atom placed in Fabry-Perot resonator. Other types of the resonators such as a waveguide [4], surface plasmon-polaritons [5], assisted by lattice vibrations of polar crystals [6] were proposed. There is much attention to exciton-plasmon coupling in particular due to a great interest in nanoplasmonics. A few papers [5, 7, 8] reported on strong exciton-plasmon coupling in both absorption and fluorescence. The splitting in both fluorescence- and absorption spectra was interpreted as Rabi splitting in the strong coupling domain. In this letter we will show that the fluorescence spectra may not show the strong coupling and analyze conditions for which a Rabi splitting may be observed. Necessary conditions will also be shown for which strong coupling between a single molecule and metal may appear.

Initially Rabi oscillations were analyzed by a quantum electrodynamics approach [1, 9]. It was also shown [10] that the Rabi splitting in absorption spectra can be simply described classically with the use of macroscopic values. A semiclassical approach [6] can be successfully applied too. In order to analyze the fluorescence spectra we will use the microscopic classical approach [11] that clearly shows the essential conditions for an observation of Rabi splitting.

Let us consider the main experimental data published in [5, 8]. Shortly: the sample consists of a layer of dye molecules deposited on a silver film. The thicknesses of the dye layer and the metal film are about 40 nm and 50 nm, respectively. The film is deposited on a glass substrate. The Kretschman configuration was used for the measurement of the reflection coefficient. The reflection data clearly show the splitting and were correctly interpreted as Rabi splitting. The fluorescence was excited by a laser beam normally incident on the air-dye layer interface and two detection schemes were used. In the first scheme, the fluorescence is detected through the glass prism; the detection was directly in the air in the second scheme. The first detector clearly showed two fluorescence peaks while the detection in the air revealed only one peak of the dye film. The lower energy peak coincides with the lower polariton branch displayed in absorption spectra and depends on the registration angle while the higher energy peak does not depend on the observation angle and coincides with the one recorded by the detector 2. The lower energy peak was interpreted [5, 8] as a manifestation of the mixed exciton-plasmon state, in other words, strong coupling. Similar fluorescence spectra with two maxima were observed [12] for a dye placed inside a low Q resonator. It should be emphasized that the observed spectra with two peaks can be explained in a simple way without any coupling effect. Indeed, the spectral dependence of the intensity of the light emitted by the molecule is determined by the product of the fluorescence spectrum of the dye layer in the absence of the metal film and the transmission spectrum of the system. Since the transmission coefficient does not reveal any resonance properties for the detector 2, the spectrum almost coincides with the one of the dye on a glass substrate. A similar feature appears for the emission through the prism at small angles. In this case the resonance properties are not manifested. However, the resonance transmission is revealed at higher values of the angles due the excitation of the surface-plasmon polariton [13]. And due to it the second peak appears. It is obvious that the position of this peak coincides with the lower branch of the surface plasmon-polariton displayed in the reflection experiment. The absence of the upper branch of the surface plasmon-polariton is caused by a asymmetry of the profile of the emitted spectra. The presence of the peak independent on the observation angle clearly indicates the absence of the strong exciton-plasmon coupling. The same Kretschman scheme was used in [7]. However, the SPP was used for the excitation of the fluorescence of a dye. The authors believed that they have observed Rabi splitting in fluorescence spectra for both detectors. Moreover, a behavior of fluorescence peak positions on the excitation angle was presented. It is obvious that this interpretation is not correct. The main reason consists in the absence of the memory between the excitation and fluorescence. The fluorescence peak position may not depend on the excitation angle. The observed behaviors seem to be of an artificial nature and are probably caused by the reabsorption of the emitted light because of using the samples with a high concentration of the dye. We will show that the strong exciton-plasmon coupling cannot be the origin of the observed fluorescence spectra with two maxima in all these publications and we will derive a necessary condition for which a strong coupling can exist.

Let us consider a radiated molecule in a vacuum as a harmonic oscillating dipole with the oscillating coordinate $z$. Then the equation of motion of the molecule placed near a body can be written as follows

$$\partial^2 z / \partial t^2 + \gamma_1 \partial z / \partial t + \omega_1^2 z = \frac{eE^r}{m} \tag{1},$$

where $m$ is the masse of the oscillating dipole, $\gamma_1$ is a decay rate, $\omega_1$ is the resonance frequency of the dipole, $e$ is the charge of the dipole, $E^r$ is the amplitude of the electric field reflected from the body. The decay rate $\gamma_1$ is caused by the radiation reaction force [14], in other words, by the action of the dipole field on itself in vacuum. A radiated spectrum is determined by a spectral behavior of $E^r$. The real part of $E^r$ is responsible for the Lamb shift, the imaginary part alters the decay rate. If the reflected field does not depend on the wavelength, for example, the reflected field from the perfect conductor, then the Lamb shift takes place [14]. Absorption of the radiated field (imaginary part of $E^r$) by a metal substrate leads to a huge increase of the decay rate [15]. However, if the reflected field has a resonance property (for example, reflected from Fabry-Perot resonator) then Eq. 1 is equivalent to the equation of the two coupled oscillators and the solution of Eq. 1 gives a splitting if the resonance frequency of the resonator coincides with the resonance frequency of the molecule. Let us show it on the example of a radiating molecule placed near a flat metal surface.

The reflected field $\vec{E}^r(\vec{r})$ is determined by $\vec{E}^r(\vec{r}) = \frac{k_0^2}{\varepsilon_0}\vec{\vec{G}}^r(\omega,\vec{r},\vec{r})\vec{p}(\vec{r})$, where $k_0$ is the wavevector in vacuum, $\varepsilon_0$ is the vacuum permittivity, $\vec{\vec{G}}^r(\omega,\vec{r},\vec{r})$ is the reflected part of the electrodynamics Green tensor, $\vec{p}(\vec{r})$ ($p(\vec{r}) = ez$) is the dipole of the radiated molecule. Looking for the solution of Eq. 1 in the form $z \sim e^{-i\omega t}$ and using $\gamma_1 = \frac{e^2 k_0^3}{6\pi\varepsilon_0 m\omega_1}$ [16], Eq. 1 can be rewritten as follows

$$-\omega^2 - i\gamma_1\omega + \omega_1^2 = \frac{\gamma_1 6\pi\omega_1 G_{zz}^r(\omega,r,r)}{k_0} \qquad (2)$$

If the distance $\Delta r$ from the radiated molecule to the surface is much less than the reduced wavelength then for the dipole oriented in the perpendicular direction the electrostatic approximation is valid and $G_{zz}^r(\omega,r,r) = \frac{1}{16\pi}\frac{\varepsilon-1}{\varepsilon+1}\frac{1}{k_0^2\Delta r^3}$ [15], where $\varepsilon$ is the dielectric function of the bulk metal. Thus, the reflected field clearly shows the resonance at

$\text{Re}\,\varepsilon = -1$. Taking the dielectric function in the form of the Drude model $\varepsilon = 1 - \frac{\omega_p^2}{\omega^2 + i\gamma_2\omega}$

one can rewrite Eq. 2 as follows

$$-\omega^2 - i\gamma_1\omega + \omega_1^2 = -\frac{\gamma_1 3\omega_1 \omega_p^2/2}{8(\omega^2 + i\gamma_2\omega - \omega_p^2/2)k_0^3\Delta r^3} \tag{3},$$

where $\omega_p$ and $\gamma_2$ are plasma frequency and damping of the plasma gas. It is seen that the reflected field in Eq. 3 has a resonance behavior at the resonance frequency $\omega_2 = \omega_p/\sqrt{2}$. Eq. 3 describes the motion of two coupled oscillators [17] and consequently has to show the splitting. At $\omega_1 = \omega_2 = \omega_0$ and near resonance Eq. 3 may be rewritten as follows

$$4\omega_0^2(\omega - \omega_0)^2 + 2i(\gamma_1 + \gamma_2)\omega_0^2(\omega - \omega_0) - \gamma_1\gamma_2\omega_0^2 - \frac{\gamma_1 3\omega_0^3}{8k_0^3\Delta r^3} = 0 \tag{4}$$

The solution gives ($\gamma_1 \ll \gamma_2$)

$$\omega \approx \omega_0 - \frac{i\gamma_2}{4} \pm \frac{\sqrt{\frac{3\gamma_1\omega_0}{2k_0^3\Delta r^3} - \gamma_2^2}}{4} \tag{5}$$

Eq. 5 clearly shows that the strong coupling is absent if $\frac{3\gamma_1\omega_0}{2k_0^3\Delta r^3} \leq \gamma_2^2$. Thus, high values of the damping or too big distances from the surface can be the reason of the absence of the strong coupling. Also for a long fluorescence time (small value of $\gamma_1$) no Rabi splitting is expected. For example, for Ag [18] $\gamma_2^2 \approx \omega_0^2/30$ while $\frac{3\gamma_1\omega_0}{2k_0^3\Delta r^3} \approx \omega_0^2/2.75$ at $\gamma_1 \approx 10^9 s^{-1}$ and $\Delta r$=0.5 nm. However, at $\Delta r$=2 nm the Rabi splitting is not expected to be resolved.

For the case of Au substrate the Drude model is not applicable and only a numerical estimation of the ratio $\frac{\varepsilon-1}{\varepsilon+1}$ can show that the solution of Eq. 2 does not exist even at 400 nm where $\varepsilon = -1$. The reason for this is the peculiar properties of the dielectric function of Au, namely its imaginary part that has a too high value.

Thus, the above analyzis clearly shows that the prerequisite of the strong coupling is the necessity of resonance properties of the reflected field initially created by the dipole. Does it take a place for the samples in [5, 7, 8]? Does the reflected field have resonance properties? It is well known [13] that the Kretschman configuration reveals resonance properties, in reflection spectra. However, the resonance frequency strongly depends on a selected angle. Such selection is realized in the reflected spectra by an angle filtration of the incident parallel beam, in other words by the phase setting along the coordinate of the surface plasmon-polariton propagation. Contrary to the reflected spectra, the emission from a molecule occurs at all angles and thus the radiation reaction signal does not display resonance properties. It means that the Green function should not reveal resonance properties. The Green function can be found elsewhere [19]. A numerically calculated absolute value of the Green function is presented in Fig. 1. Following [5] we have taken the thickness of the Ag film to be 50 nm, the thickness of the layer with the dye to be 40 nm (a) and 1 nm (b). The refraction index of the matrix with the dye and the glass prism are assumed as 1.6, and 1.5, respectively. Some contribution to the dielectric function of the layer with dye was modeled by the Lorentz formula with the resonance at 2099 mEv. The source of the fluorescence was placed in the centre of the film with the dye. In other words, the distance from the dipole to the metal film was 20 nm (a) and 0.5 nm (b). Fig. 1a shows some absorption by the dyes at 2099 mEv, however, no resonance enhancement at any frequency. Conversely, Fig. 1b clearly demonstrates the resonance at the energy 3356 mEv. It corresponds to the wavelength 370 nm. The absolute value of the real part of the dielectric function of the silver at this wavelength is 2.58 approximately equaled to $2.56=1.6^2$. Thus, this resonance was described by the above described equations. Instead of the resonance condition $\operatorname{Re}\varepsilon = -1$ for the molecule in vacuum, the resonance is revealed at $\operatorname{Re}\varepsilon = -\varepsilon_1$ if the molecule is placed in the medium with the dielectric function $\varepsilon_1$. Thus, the resonance can be revealed if the distance from the molecule to metal is about 1 nm or less.

In the case of the absorption spectra the Rabi splitting is also determined by the reradiated reflection field. However, contrary to the fluorescence the absorption takes place for many molecules, the excitation phases of which are determined by the incident beam. Due to it the reradiated reflection field is selected for the angle that equals to the incident one. As we have mentioned already the Kretschmann configuration reveals the resonance properties at the fixed angles. Note that Rabi splitting would not reveal at the absorption of the single molecule.

If the molecule is placed near a sphere and the distance to its surface is much less than the radius, the reflected field is determined [20] by the same equation as for the case of the flat surface. In other words, at $\operatorname{Re}\varepsilon = -1$. If the distance between the molecule and spherical particle is larger than the radius of curvature, the reflected field is proportional to $1/(\varepsilon + 2)$, however, it is simply to show that the amplitude is much less than for the small distances and the strong coupling may not be revealed at $\operatorname{Re}\varepsilon = -2$. In other words, the Rabi splitting for the system single molecule - spherical particle may be observed also only in UV. The reflected radiation field for the case of a molecule near a spheroidal particle is determined [21] by the same equation as for the spherical particle. So, the resonance in the reflected field is at $\mathrm{Re}\,\varepsilon = -1$ too. It was recently mentioned [22] that for the coupled spherical particles the resonance condition has a red shift and can be resolved at 450 nm. However, such a supposition has to be verified by the calculation of the Green function.

In conclusion, necessary conditions that the strong exciton-plasmon coupling can be observed in fluorescence spectra are a frequency where $\mathrm{Re}\,\varepsilon = -1$ and a distance between the molecule and metal surface of about 1 nm or less. Only Ag can display the Rabi splitting in the fluorescence in the UV spectral domain.

This work would not appear without a stimulant discussion initiated by Ulrich Fischer. I am very grateful him also for editing English. Author acknowledges the Deutsche Forschungsgemeinschaft (German Research Foundation) and STMS for financial support.

Figures captures.

Fig. 1. Spectral behavior of the absolute value of the zz-component of the Green function at the thickness of layer with the dye of 40 nm (a) and 1 nm (b).

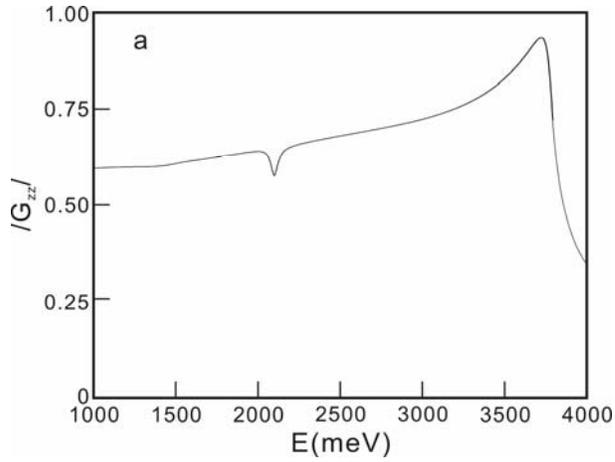

Fig. 1a.

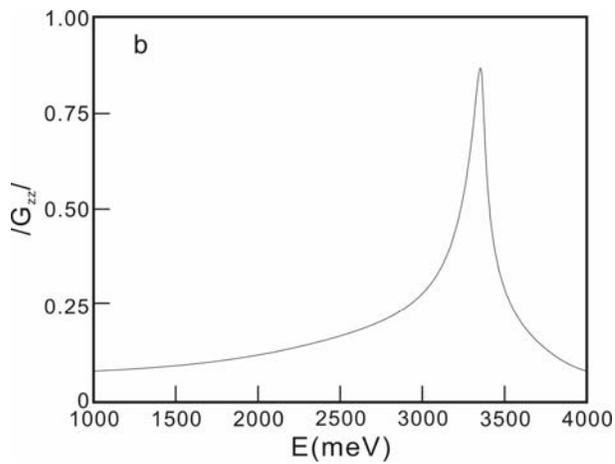

Fig. 1b.